\documentclass[aps,apl,twocolumn,superscriptaddress]{revtex4}

\usepackage{graphicx}
\usepackage{epstopdf}
\usepackage{comment}
\usepackage{amsmath,amssymb}
\usepackage{bm}
\usepackage{float}
\usepackage[utf8]{inputenc}
\usepackage[T1]{fontenc}

\newcommand{\ctext}[1]{\raise0.2ex\hbox{\textcircled{\scriptsize{#1}}}}
\newcommand{\vect}[1]{{\mbox{\boldmath $#1$}}}

\begin{document}

\title{On-chip magnon polaron generation in mode-matched cavity magnomechanics}

\author{Daiki Hatanaka}
\email[]{daiki.hatanaka@ntt.com}
\thanks{These authors contributed equally to this work}
\affiliation{Basic Research Laboratories, NTT, Inc., Kanagawa, Japan}

\author{Motoki Asano}
\email[]{motoki.asano@ntt.com}
\thanks{These authors contributed equally to this work}
\affiliation{Basic Research Laboratories, NTT, Inc., Kanagawa, Japan}

\author{Megumi Kurosu}
\affiliation{Basic Research Laboratories, NTT, Inc., Kanagawa, Japan}

\author{Yoshitaka Taniyasu}
\affiliation{Basic Research Laboratories, NTT, Inc., Kanagawa, Japan}

\author{Hajime Okamoto}
\affiliation{Basic Research Laboratories, NTT, Inc., Kanagawa, Japan}

\author{Hiroshi Yamaguchi}
\affiliation{Basic Research Laboratories, NTT, Inc., Kanagawa, Japan}

\date{\today}

\begin{abstract}
Generation of magnon polarons, which are hybridized states resulting from strong magnon-phonon coupling, is a key to enabling coherent manipulation in acoustic and spintronic devices. However, the conventional device configuration, a magnetic thin film on a thick piezoelectric layer, often has difficulty achieving a large magnon-phonon coupling due to a very small spatial mode overlap. Here, we demonstrate generation of magnon polarons by using a mode-matched on-chip magnomechanical system. A configuration with a thin piezoelectric film on a magnetic layer several micrometers thick was found to sustain deeply distributed magnon modes that enable magnetoelastic coupling to phonons over almost the entire mode volume. The enhanced spatial mode overlap generated magnon polarons whose spectra showed distinct avoided crossing. This magnomechanical system will facilitate utilization of coherent magnon-phonon conversion and their hybrid states in functional phononic devices.
\end{abstract}

\maketitle

\begin{figure*}[t]
	\begin{center}
		\vspace{-0.cm}\hspace{-0.0cm}
		\includegraphics[width=18cm]{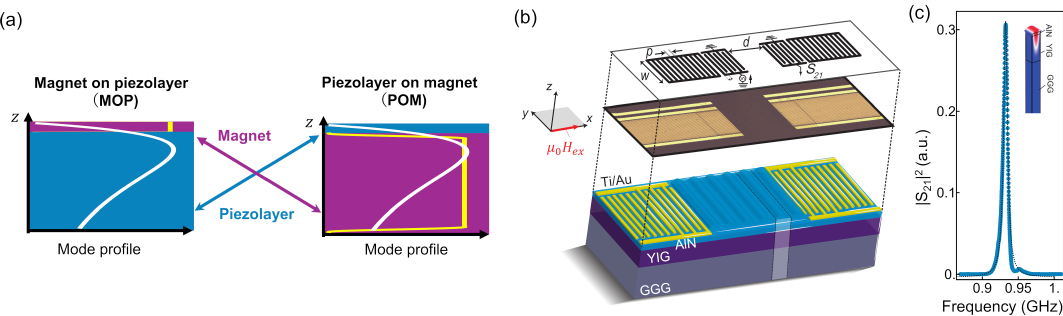}
		\vspace{0.cm}
		\caption{(a) Schematic mode profiles of the phonon mode (the spatial profile of shear strain, $\epsilon_{\rm zx}$) in a conventional (left) and our magnomechanical system (right). In the conventional system, a magnet (purple) is placed on the piezoelectric layer (blue), and the magnon mode volume (yellow solid line)  is localized in a much narrower region compared with the phonon modes. On the other hand, our system has a thick magnet below the piezoelectric layer, so the magnon and phonon modes mostly overlap. The mode profile of the shear strain ($\epsilon_{zx}$) is denoted by the white solid lines . (b) Bottom: Schematic view of a cavity magnomechanical system containing YIG ($L=5.7$ $\mu$m) on GGG substrate, covered by piezoelectric AlN film (0.4 $\mu$m). SAWs are generated and detected piezoelectrically through inter-digit transducers (IDTs; highlighted in yellow). An external magnetic field is applied parallel to the SAW propagation direction (red solid arrow) ($x$-axis). Middle: An optical microscope image of a SAW resonator (cavity), where the IDTs are sandwiched by Bragg reflectors. Top: Geometry of the metallic array structures whose period, aperture size, and IDT distance are $p=$ 2.0 $\mu$m, $w=$ 190 $\mu$m and $d =$ 300 $\mu$m respectively.  (c) Spectral response of a SAW cavity where acoustic resonance occurs at $f_a=$ 0.93 GHz, excited and measured through IDTs. A mode profile of the displacement amplitude in part of the SAW cavity, simulated with the finite element method (FEM), is shown on the right.}
		\label{fig 1}
		\vspace{-0cm}
	\end{center}
\end{figure*}

\begin{figure}[t]
	\begin{center}
		\vspace{-0.cm}\hspace{-0.0cm}
		\includegraphics[width=8cm]{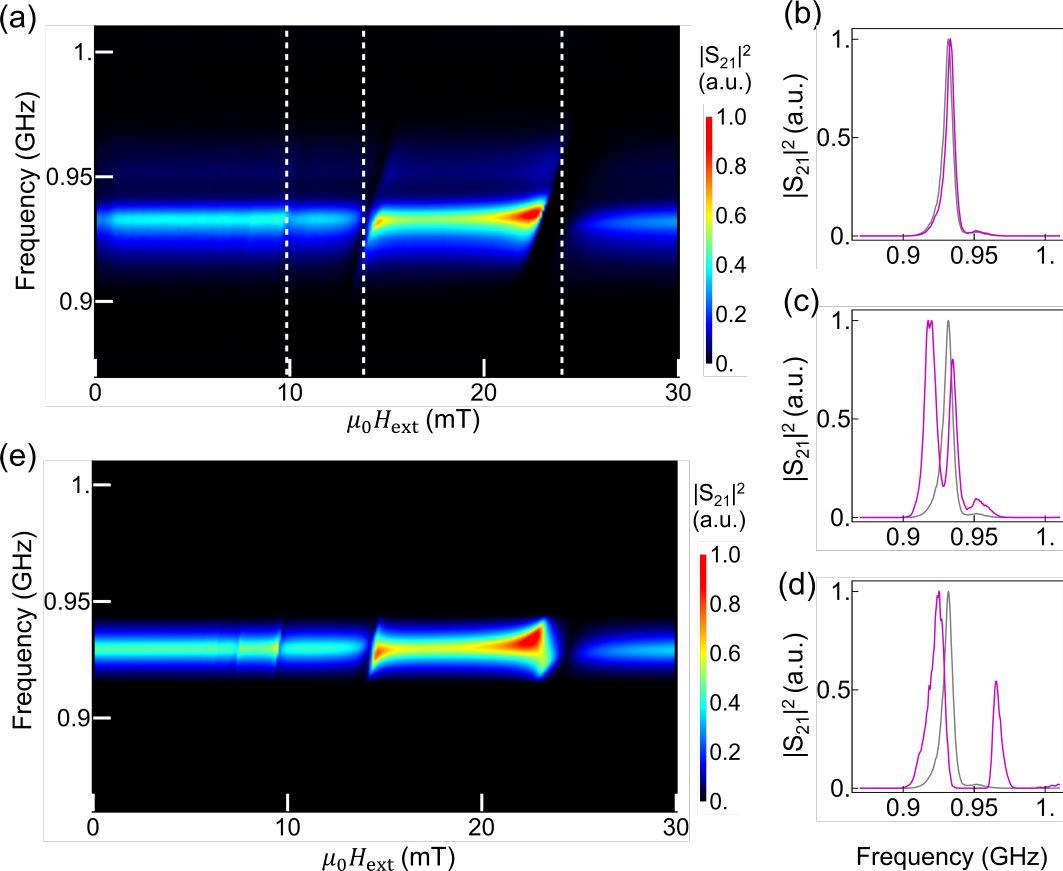}
		\vspace{0.cm}
		\caption{(a) Magnetoelastic response of SAW cavity. The dashed lines correspond to the external magnetic field showing a large frequency modulation. (b)-(d) Characteristic phonon spectra (the purple lines) at (b) $\mu_0H^{(0)}_\mathrm{res}=24.2$ mT, (c) $\mu_0H^{(1)}_\mathrm{res}=13.8$ mT, and (d) $\mu_0H^{(2)}_\mathrm{res}=10.0$ mT. The gray lines shows the phonon spectra at $\mu_0H_\mathrm{ext}=0$ mT. (e) Numerically calculated phonon spectra with respect to the external magnetic field $\mu H_\mathrm{ext}$ via magnon-phonon coupling.}
		\label{fig 2}
		\vspace{-0cm}
	\end{center}
\end{figure}

\begin{figure}[t]
	\begin{center}
		\vspace{-0cm}\hspace{-0.0cm}
		\includegraphics[width=8cm]{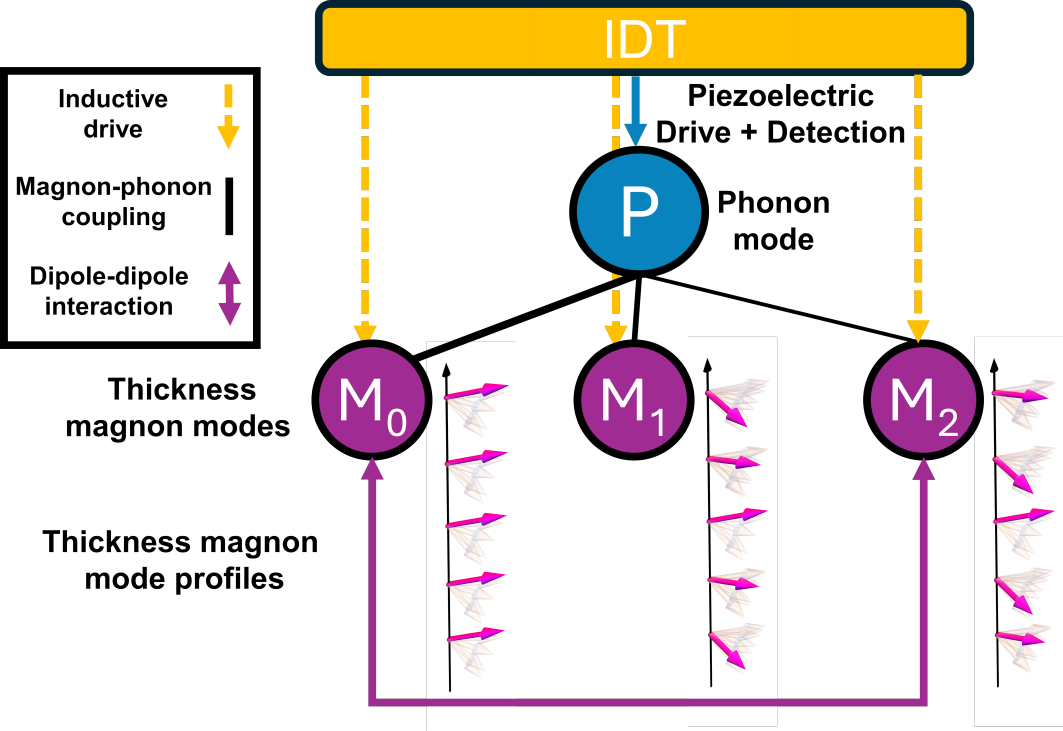}
		\vspace{0.cm}
		\caption{Theoretical model of the mode-matched magnomechanical system. The inset vector profiles show the thickness modes along the thickness ($z$) and propagation ($x$) directions. The mode profiles $\psi_\mathrm{M}(z)$ are also explicitly described.}
		\label{fig 3}
		\vspace{-0cm}
	\end{center}
\end{figure}

Magnomechanics is an emerging research field that intersects phononics and magnonics, where different energy states, spin motions and mechanical vibrations, couple to give rise to enriched hybridization physics \cite{li2021advances, liao2024hybrid}. Here, phononics, particularly on micro and nanodevice platforms such as surface acoustic waves (SAWs) and phononic crystals, has developed sophisticated capabilities for on-chip excitation, guiding, and detection of phonons \cite{benchabane_pnc2, fu2019phononic, activeSAW_loncar, hatanaka_hyperPnC, yao2025perspectives}. Its integration with magnonics allows not only phonon transport to be magnetically modulated and controlled but also nontrivial phonon dynamics by introducing distinctive magnonic features such as nonlinearity, chirality, and nonreciprocity \cite{weiler_adfmr, sasaki2017nonreciprocal, kraimia2020time, xu2020nonreciprocal, tateno2021einstein, hatanaka2023phononic, kuss2023giant}. A prerequisite for realizing a magnomechanical system is the generation of magnon polarons through a strong magnon-phonon coupling in which the coupling rate, $g_{\mathrm{S}}$, is larger than the magnon and phonon dissipation rates ($\kappa_\mathrm{M}$ and $\kappa _\mathrm{A}$, respectively). Stable generation of magnon polarons may further lead to various novel phononic phenomena involving magnon Bose-Einstein condensation \cite{demokritov2006bose, bozhko2016supercurrent, bozhko2019bogoliubov} and alternative device technologies \cite{flebus2017magnon, hayashi2018spin, godejohann2020magnon}.

The key parameter for realizing hybrid states is the magnon-phonon coupling strength, which is predicted to be proportional to the magnon-phonon mode overlap ($V_{\rm OV}$) and the square root of the resonance frequency ($\omega_{\rm A}$), i.e., $g_{\rm S} \propto V_{\rm OV} \sqrt{\omega_{A}}$ \cite{hatanaka_sawmag, asano2023cavity, matsumoto2024magnon}. In the layer structures used in magnomechanical studies, a magnetic thin film is deposited on a thick piezoelectric film. In this magnet on piezolayer (MOP) configuration, SAWs interact with the magnetic thin film only near its surface [see the left in Fig. 1(a)]. This is because the magnon is confined within the magnetic thin film, in contrast to the acoustic mode, which can easily extend into the underlying bulk material of the substrate. Therefore, the interaction occurs only in a small portion of the overall acoustic mode volume ($V_{\rm A}$), limiting the magnon-phonon mode overlap ($V_{\rm OV} \ll V_{\rm A}$) \cite{hatanaka_sawmag, asano2023cavity, matsumoto2024magnon, hwang2024strongly}. To date, strong coupling in magnon-SAW systems has been experimentally demonstrated only in a CoFeB-SAW hybrid system \cite{hwang2024strongly}. The employed layer structure had an MOP configuration, i.e. a magnetic film dozens of nanometers thick on a piezoelectric layer, inducing only a slight mode overlap of spin waves with the SAW cavity mode. To compensate for the reduction in magnon-phonon mode overlap, it was required to increase the acoustic resonance frequency to $\omega_{\rm A}/(2 \pi)=$ 6.5 GHz, which is higher than that commonly used in standard SAW technologies.

Here, we demonstrated strong magnon-phonon coupling and magnomechanically-driven magnon polarons at a reduced SAW frequency of 0.93 GHz by employing an alternative layer structure in which a thin piezoelectric film was deposited on a thick magnetic layer (POM configuration). By placing yttrium iron garnet (YIG) with an ultra-low $\kappa_{\rm M}$ under a piezoelectric AlN layer, the thickness of the magnetic film could be on the same length scale as the penetration depth of the SAWs. Thanks to this mode-matched configuration, the thick magnetic film enabled a large spatial mode overlap between the magnon and phonon modes [see right in Fig. 1(a)]. Thus, the magnomechanical system sustained coherent magnon polarons, i.e. magnetoelastic waves, at the moderate excitation frequency.

The key material of our device is YIG film that is epitaxially grown on a gadolinium gallium garnet (GGG) substrate and is covered by an aluminum nitride (AlN) film, as shown in Fig. 1(b). The SAW cavity consists of two inter-digit transducers (IDTs) surrounded by two metallic Bragg reflectors and is formed on the AlN/YIG/GGG substrate. Piezoelectric transduction at the IDTs for excitation and detection of SAWs is enabled by the AlN film. The propagating SAWs with a wavelength of $\lambda_A = 4.0$ $\mu$m are reflected by the Bragg lattices, wherein the reflected waves constructively interfere with the incident waves. The SAW resonance can be observed by measuring $S_{21}$ spectra via the IDTs [see Fig. 1(c)]; the measured resonance frequency was $\omega_\mathrm{A} / (2\pi)=$ 0.93 GHz and the spectral linewidth was $\kappa_\mathrm{A}/(2\pi)=7.75$ MHz (i.e., $Q_\mathrm{A}=\omega_\mathrm{A}/\kappa_\mathrm{A}=120$). The magnetic YIG layer has a thickness of $L=5.7$ $\mu$m and is located under the AlN layer in our POM configuration. The SAW strain is mostly concentrated in the YIG layer so that the magnetic and acoustic cavities share nearly the same spatial region whose sufficiently large $V_{\rm OV}$ ($\approx V_{\rm A}$) leads to strong magnon-phonon coupling.

To experimentally confirm the enhanced coupling in the POM configuration, the phonon spectrum was measured while changing the strength of the dc magnetic field, $\mu_0 H_\mathrm{ext}$ [see Fig. 2(a)]. Here, the dc magnetic field was externally applied in the plane of the substrate surface and parallel to the propagation direction of the SAWs ($\phi_h=0^\circ$). This setup was employed because the coupling through the shear strain component, which is strongly enhanced in this mode-matching configuration, is maximized in this field direction. In addition, it minimizes the dipole-dipole interaction, which leads to a spectrally isolated magnon mode especially at high $\mu_0 H_\mathrm{h}$ region \cite{chumak2015magnon, pirro2021advances}, so that the isolated spectrum allows for a quantitative comparison of the measured spectra with theoretical calculations. The measured SAW resonant spectrum was strongly modulated around $\mu_0 H^{(0)}_\mathrm{res}=24.2$ mT, $\mu_0 H^{(1)}_\mathrm{res}=13.8$ mT, and $\mu_0 H^{(2)}_\mathrm{res}=10.0$ mT [see Fig. 2(b), (c), and (d), respectively]. A small frequency shift was observed at $\mu_0 H^{(2)}_\mathrm{res}$, whereas avoided crossings, which are signatures of the magnon-phonon strong coupling, appeared at $\mu_0 H^{(1)}_\mathrm{res}$ and $\mu_0 H^{(0)}_\mathrm{res}$.

The measured spectrum has two unusual features that are not observed in a simple system with mode-mode strong coupling. One is that the frequency modulation occurred at the multiple external magnetic field values ($\mu H^{(0)}_{\rm res}$, $\mu H^{(1)}_{\rm res}$, $\mu H^{(2)}_{\rm res}$). This feature suggests that multiple magnon modes are coupled to one phonon mode. The other is that the phonon amplitude is asymmetric around the points of avoided crossing. The amplitude at the lower (higher) edge of the crossing field, $\mu H^{(0)}_{\rm res}$, is enhanced (suppressed), whereas the opposite feature is observed around $\mu H^{(1)}_{\rm res}$. This second feature is clearly different from the typical spectrum of avoided crossing, where the amplitude distribution is nearly symmetric across the coupling point. To theoretically reproduce these features and quantify the strength of the magnon-phonon coupling, we modeled the magnon-phonon coupling assuming a single phonon mode coupled to multiple magnon modes (a.k.a. thickness modes) confined within the relatively thick magnetic film ($kL > 1$) \cite{pirro2021advances}. Here, each magnon mode has a different spatial profile along the thickness ($z$) direction, $\psi_{\mathrm{M},l}(z)=\cos(l\pi z/L)$, where $l$ is the mode number. Thus, the mode overlap integral $V_\mathrm{OV}$ is expected to depend on the mode number $l$, leading to a mode-dependent magnon-phonon coupling constant. We also assumed that the magnon is not only excited by acoustic phonons but also by the inductive radio-frequency field directly generated from the IDT electrodes. As shown below, the second assumption well explains the asymmetric feature around the avoided crossing.

The coupled mode equation can be formulated by combining the equation of elastic motion and the Landau–Lifshitz–Gilbert (LLG) equation including exchange and dipole-dipole interactions, as follows:
\begin{align}
\dot{\vect{m}_l}=D_l\vect{m}_l+\sum_{k\neq l}R_{k}\vect{m}_k+ig_{\mathrm{MA},l}(0,1)^T b+\vect{m}_{\mathrm{in},l},\\
\dot{b}=\left(i\Delta-\frac{\kappa_A}{2}\right)b+ig_{\mathrm{AM},l} (0,1)^T\vect{m}_{l}+b_\mathrm{in}.
\end{align}
where $\vect{m}_l$ is a two-dimensional vector showing magnon modes in the $l$-th thickness mode, and $b$ is the amplitude of the phonon mode. The magnon dynamics are governed by the intramode exchange interaction given by the 2$\times$2 matrix, $D_l$, the intermode dipole-dipole interaction given by the 2$\times$2 matrix, $R_k$, the magnon-phonon coupling rate, $g_\mathrm{MA}$, and the direct magnon excitation through inductive driving via the IDT electrodes, $\vect{m}_{\mathrm{in},l}$. In the same manner, the phonon dynamics with the finite detuning $\Delta$ and dissipation rate $\kappa_\mathrm{A}$ are governed by the magnon-phonon coupling rate, $g_\mathrm{AM}$, and the direct phonon excitation through the piezoelectric driving force, $b_\mathrm{in}$, via the IDTs. The theoretical model is illustrated in Fig. 3, and a detailed formulation is given in the Supplemental Information.

Figure 2(e) shows the numerically calculated phonon spectra using the coupled mode equation. Here, several unknown and structure-dependent parameters were estimated via Monte-Carlo simulations (see the Supplemental Information). The numerical results are in good agreement with the experimental results in Fig. 2(a). In the following sections, we discuss two features in our mode-matched magnomechanical system based on the theoretical calculations.

\begin{figure*}[tbh]
	\begin{center}
		\vspace{-0.5cm}\hspace{-0.0cm}
		\includegraphics[scale=0.8]{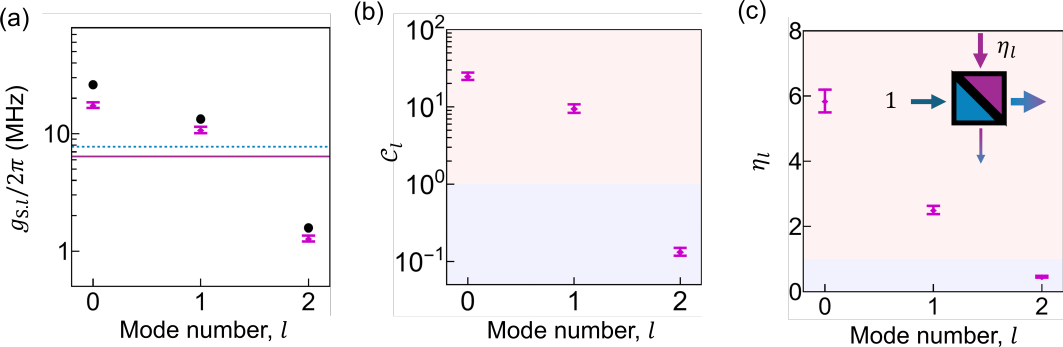}
		\vspace{0.5cm}
		\caption{(a) Symmetrized magnon-phonon coupling constant with respect to the magnon mode numbers. The purple diamonds are numerically estimated values from the experimental results. The black dots are values fully calculated from the analytical expressions. The phonon and magnon dissipation rates are denoted by the dotted blue line and the solid purple line, respectively. (b) Magnon-phonon cooperativity with respect to the magnon mode numbers. (c) Dimensionless excitation ratio with respect to the magnon mode number. Each error bar shows the estimated standard deviation in the Monte-Carlo simulation for the parameter optimization.}
		\label{fig 4}
		\vspace{-0cm}
	\end{center}
\end{figure*}

First, we can conclude that strongly hybridized magnon polarons are generated by the large mode overlap between the magnons and phonons. This conclusion is supported by two experimental findings, i.e., symmetrized magnon-phonon coupling constants ($g_{\mathrm{S},l}\equiv\sqrt{g_{\mathrm{AM},l}g_{\mathrm{MA},l}}$) and cooperativity ($C_l$). Figure 4(a) shows $g_{\mathrm{S},l}$ in each thickness mode estimated from the experimental results via numerical parameter optimization (the purple diamonds) and the fully analytical results with the parameter values from the literature (the black solid circle). Apparently, these values, especially for the fundamental ($l = 0$) and the first mode ($l = 1$), show coupling constants much larger than the phonon dissipation rate (the dotted blue line) and the magnon dissipation rate (the solid purple line), i.e., $\kappa_{\rm A}, \kappa_{\rm M} < g_{\mathrm{S},l}$. Another parameter for evaluating the quality of magnon polarons is magnon-phonon cooperativity, $\mathcal{C}_l=4g^2_{\mathrm{S},l}/(\kappa_\mathrm{A}\kappa_\mathrm{M})$ for each mode number, $l$ [see Fig. 4(b)]. The maximum cooperativity $\mathcal{C}_0= 25.0 \pm 5.5$ is obtained in the fundamental mode. This value is much larger than the unity that was reported in conventional magnomechanical systems \cite{hatanaka_sawmag, matsumoto2024magnon}, and it indicates the formation of coherent magnon polarons in our device. The commonly observed mode dependence, i.e. a longer wavelength magnon mode leads to a larger coupling, clearly proves that the mode-matched thick piezoelectric layer in the MOP configuration is advantageous for enlarging the coupling constant. 

Second, asymmetric magnon polarons, i.e., the non-uniform phonon spectra around $\mu H^{(0)}_{\rm res}$ and $\mu H^{(1)}_{\rm res}$, were generated via magnon-phonon interference. This can be qualitatively interpreted in terms of a balanced beam splitter model with two different input amplitudes [see the inset of Fig. 4(c)]. As the strong coupling induces the polaron mode, the peak amplitudes of the phonon modes with avoided crossing become uniform when only the phonon mode is excited (i.e., $\eta_l=0$). However, when the phonon and magnon modes are simultaneously excited, the generated polaron mode shows asymmetric amplitudes due to magnon-phonon interference. Their simulated phonon spectra in comparison with the symmetric and asymmetric magnon polarons are shown in the Supplementary Information. From the experimental results and numerical analysis, we can quantify the magnon amplitude via inductive driving by the dimensionless excitation ratio, $\eta_l\equiv \sqrt{g_{\mathrm{AM},l}/g_{\mathrm{MA},l}}|\vect{m}_{\mathrm{in},l}|/b_\mathrm{in}$. The dimensionless excitation ratio exceeds unity ($\eta_l>1$) in modes $l=0$ and 1, which means that the hybridized mode is primarily driven by the inductively-induced magnon mode rather than the phonon one. This indicates that ours is a new approach to driving a magnomechanical system that may enable advanced manipulation technology using parametric excitation via nonilnear magnetic interaction \cite{makiuchi2024persistent} and the spin Seebeck effect \cite{harii2019spin, uchida2011long}.  

Finally, we comment on the intermode dipole-dipole interaction, denoted by $R_k$ in Eq. (1), which allows coupling between both odd (or even) thickness magnon modes. The interaction leads to indirect coupling between the second magnon mode and the phonon mode through the zeroth-order fundamental magnon mode. This is the origin of the phonon spectral modulation around $\mu_0 H^{(2)}_\mathrm{res}\sim $ 10.0 mT, where the second-order weak magnon-phonon coupling in the second thickness mode was observed. We believe that further experiments at $\phi_\mathrm{h}\neq 0$ will pave the way to controlling magnon polarons whereby the multiple magnon modes couple to a single phonon via magnon-phonon and dipole-dipole couplings that have a scalable multimode architecture. 

In conclusion, we demonstrated magnon polaron generation by utilizing a mode-matched magnomechanical system. A well-engineered large spatial mode overlap between the phonon mode and multiple magnon modes was achieved, and the unusual spectral behavior could be described both analytically and in terms of a numerical optimization. The magnomechanical platform proposed in this study provides an appropriate support for magnon polarons on a tiny planar system and is a promising technology for diverse controls with potential application to microwave information processing.

This work was supported by JSPS KAKENHI Grants, Numbers JP21H05020, JP23H05463 and JP24H02235. We are grateful for the stimulating discussions at the meeting of the Cooperative Research Project of the Research Institute of Electrical Communication, Tohoku University.

\bibliography{00_reference.bib}

\clearpage
\pagebreak
\widetext
\begin{center}
\textbf{\large Supplemental information}
\end{center}
\setcounter{equation}{0}
\setcounter{figure}{0}
\setcounter{table}{0}
\setcounter{page}{1}
\makeatletter
\renewcommand{\theequation}{S\arabic{equation}}
\renewcommand{\thefigure}{S\arabic{figure}}
\renewcommand{\bibnumfmt}[1]{[S#1]}
\renewcommand{\citenumfont}[1]{S#1}

\section{Theory of magnon-phonon coupling}
\subsection{Model overview}
Here, we give an overview of the coupled magnon-phonon model in order to interpret our experimental results. The equations of motion for magnon modes and phonon modes take the form of coupled mode equations ,
\begin{align}
\dot{\vect{m}_l}(t)=D_l\vect{m}_l(t)+\sum_{k\neq l}R_{k}\vect{m}_k+ig_{\mathrm{MA},l}(0,1)^Tb(t)+\vect{m}_{\mathrm{in},l},\\
\dot{b}(t)=\left(i\Delta-\frac{\kappa_A}{2}\right)b(t)+i\sum_{j}g_{\mathrm{AM},j} m_{j,1}(t)+b_\mathrm{in}.
\end{align}
The first equation, i.e., the equation of motion for the $l$th magnon thickness mode ($\vect{m}_l$), includes the exchange interaction terms in the 2x2 matrix, $D_l$, and the dipole-dipole interaction in the 2x2 matrix, $R_k$ \cite{kalinikos1986theory}. The magnon-phonon coupling via magnetostriction is introduced through the third term with the coupling rate, $g_{\mathrm{MA},l}$. The fourth term is the inductive driving of the magnon modes via the external radio-frequency (rf) field. 

The second equation, i.e., the equation of motion for the phonon mode ($b$), includes the detuning, $\Delta$, and the damping factor, $\kappa_\mathrm{A}$, and the magnetostrictive coupling term with $g_{\mathrm{AM},l}$. Here, the symmetrized magnon-phonon coupling constant, $g_\mathrm{S}$, is defined as
\begin{align}
g_{\mathrm{S},l}\equiv\sqrt{g_{\mathrm{AM},l}g_{\mathrm{MA},l}}=&b_2k \cos\phi_M \sqrt{\frac{\gamma M_{s}}{2\omega_A\rho_0}}\frac{V_{\mathrm{ov},l}}{\sqrt{V_A V_{\mathrm{M},l}}}.
\end{align}
Since this value has the dimension of frequency, the condition for the magnon polaron, i.e., strong magnon-phonon coupling, is given by $g_\mathrm{S}\gg \kappa_\mathrm{A},\kappa_\mathrm{M}$. The magnon-phonon coupling completely originates from the shear strain field via magnetostriction since $\phi_\mathrm{ext}=0$ is set. A detailed description of the derivation of the above coupled mode equation is given below.

\subsection{Magnon model}
The magnon in a thick ferromagnetic film, $\vect{m}(\vect{r},t)$, is decomposed into orthogonal modes along the thickness direction, $z$:
\begin{align}
\vect{m}(\vect{r},t)&=\sum_l \psi_{M,l}(z)e^{-i\omega_l t+ik_{l}(\cos\phi_S x+\sin\phi_Sy)}(m_{l,1}(t)\vect{e}_{1}+m_{2}(t)\vect{e}_{l,2})\nonumber\\
&\equiv\sum_{l} \psi_{M,l}(z)e^{-i\omega_l t+ik_{l}(\cos\phi_S x+\sin\phi_Sy)}\vect{m}_l(t)
\end{align}
where $\omega_l$ and $k_l$ correspond to the angular frequency and wavevector of the $l$th magnon modes (a.k.a. thickness modes). Here, the magnon modes propagate along the wavevector of $\vect{k}_l=(k_l\cos\phi_S,k_l\sin\phi_S)$. The external magnetic field is exerted along an arbitrary vector in the $x-y$ plane. The magnetic precession can be described in the laboratory frame, $(m_x,m_y,m_z)$, and in the magnetization frame, $(m_1,m_2,m_3)$; the coordinates are related as
\begin{align}
m_x=&-m_2 \sin\phi_M +m_3\cos\phi_M,\\
m_y=&m_2\cos\phi_M +m_3\sin\phi_M,\\
m_z=&-m_1,
\end{align}
where $\phi_M$ is the magnetization angle.

The dynamics of the magnon precession of the $l$th magnon mode, $\vect{M}_l$, which is a two-dimensional vector, are given by the LLG equation including the exchange and dipole-dipole interactions. It can be formulated with superoperators, $\mathcal{L}_j$, as follows:
\begin{align}
    \dot{\vect{m}}_l(t)=(\mathcal{L}_0+\mathcal{L}_\mathrm{D}+\mathcal{L}_\mathrm{MM}+\mathcal{L}_\mathrm{drive})\vect{m}_l(t),\label{mag_form}
\end{align}
where $\mathcal{L}_0$ denotes both the Zeeman effect and exchange interaction, $\mathcal{L}_\mathrm{D}$ denotes the dipole-dipole interaction, $\mathcal{L}_\mathrm{MM}$ denotes the magnomechanical interaction via magneto-rotation coupling, and $\mathcal{L}_\mathrm{drive}$ denotes inductive driving of the magnon modes. The third term reduces to 
\begin{align}
    \mathcal{L}_\mathrm{MM}\vect{m}_l=-\gamma \vect{m}_l\times\mu_0\vect{H}_\mathrm{eff,MM}
\end{align}
where the effective magnetic field from the magnetostriction is given by
\begin{align}
  \mu_0\vect{H}_\mathrm{eff,MM}=&-M^{-1}_s \sum_{j=1,2}\partial_{m_j}\mathcal{E}_\mathrm{MM} \vect{e}_j, \nonumber\\
  \sim&- \sum_{j=1,2}\partial_{m_j}\left[-b_1m_2\sin2\phi_M\varepsilon_{xx}-b_2m_1\cos\phi_M\right] \vect{e}_j, \nonumber\\
  =&b_2 \cos\phi_M \varepsilon_{xz}\vect{e}_1 +b_1 \sin2\phi_M \varepsilon_{xx}\vect{e}_2.
\end{align}
In particular, when the external magnetic field is exerted along the $x$ direction, $\phi_\mathrm{ext}=0$, it can be approximated to
\begin{align}
\mathcal{L}_\mathrm{MM}\vect{m}_l\sim&2b_2\cos\phi_M\varepsilon_{xz}\vect{e}_2,\nonumber\\
    \sim&b_2\cos\phi_M\partial_x \psi_{P,z}(\vect{r})\vect{e}_2U(t),\nonumber\\
    \equiv&iG_{\mathrm{MA},l}(0,1)^T b(t)
\end{align}
where $G_\mathrm{MA}=b_kk\gamma$. Here, the second approximation originating from the spatial profile of the phonon mode, $\psi_{P,z}(\vect{r})$, satisfies $\partial_x \psi_{P,z}\sim k\psi _{P,z}\gg\partial_j\psi_{P,z}$ $(j=y,z)$. 

The first and second term in Eq. (\ref{mag_form}) were formulated for thick ferromagnetic film \cite{kalinikos1986theory}. The first term is a 2x2 matrix $D_l\vect{m}_l$ showing the intramode contribution from the Zeeman and exchange interactions. Moreover, the second term is a 2x2 matrix $R_k\vect{m}_k$ $(k\neq l)$ showing the intermode contribution from the dipole-dipole interaction. Moreover, linear inductive driving is assumed so that $\mathcal{L}_\mathrm{drive}\vect{m}_l\sim \tilde{\vect{m}}_{\mathrm{in},l}$.

By separating the spatial and temporal parts as $\vect{m}_l\to \vect{m}_l(t)\psi_\mathrm{M}(\vect{r})$, $\vect{m}_{\mathrm{in},l}\to\vect{m}_{\mathrm{in},l}\psi_\mathrm{M,in}(\vect{r})$ and $b\to b(t)\psi_\mathrm{A}(\vect{r})$ and taking the integral while multiplying $\psi_\mathrm{M}(\vect{r})$ on both sides, we obtain the following equation of motion,
\begin{align}
\dot{\vect{m}_l}(t)=D_l\vect{m}_l(t)+\sum_{k\neq l}R_{k}\vect{m}_k+ig_{\mathrm{MA},l}(0,1)^T b(t)+\vect{m}_{\mathrm{in},l},
\end{align}
where
\begin{align}
g_\mathrm{MA}=&G_\mathrm{MA} \frac{\int\mathrm{d}^3\vect{r} \psi_{\mathrm{M},l}(\vect{r})\psi_\mathrm{A}(\vect{r})}{\int\mathrm{d}^3\vect{r} \psi^2_{\mathrm{M},l}(\vect{r})}\equiv b_2k\gamma \frac{V_\mathrm{OV}}{V_\mathrm{M}},\\
\vect{m}_{\mathrm{in},l}=&\frac{\int\mathrm{d}^3\vect{r} \psi_{\mathrm{M},l}(\vect{r})\psi_\mathrm{M,in}(\vect{r})}{\int\mathrm{d}^3\vect{r} \psi^2_{\mathrm{M},l}(\vect{r})}\tilde{\vect{m}}_{\mathrm{in},l}(t).
\end{align}

\subsection{Acoustic model}
The equation of motion for the displacement in the acoustic resonator is given by
\begin{align}
\ddot{U}_a+\kappa_A \dot{U}_a+\omega^2_A U_a=\frac{\int\mathrm{d}^3\vect{r}\vect{F}_\mathrm{MM}\cdot\vect{\psi}_\mathrm{A}(\vect{r})}{\rho \int \mathrm{d}^3\vect{r} \vect{\psi}_A(\vect{r})\cdot\vect{\psi}_\mathrm{A}(\vect{r})}+f_\mathrm{dr},
\end{align}
where $U_a$ is the dynamical part of the acoustic mode, and $\vect{\psi}_\mathrm{A}(\vect{r})$ is the spatial distribution of the acoustic mode. The magnetostrictive force, $\vect{F}_\mathrm{MM}$, is given by
\begin{align}
[\vect{F}_\mathrm{MM}]_\mu=&\sum_{j={X,Y,Z}}\partial_j\partial_{\varepsilon_{\mu j}}\mathcal{E}_\mathrm{MM},
\end{align}
where $[\vect{v}]_\mu$ denotes the $\mu$-component of the vector $\vect{v}$. The energy density, $\mathcal{E}_\mathrm{MM}$, is given by
\begin{align}
\mathcal{E}_\mathrm{MM}=&\mu_0 M_s \sum_{j,k={X,Y,Z}} b_{jk} \varepsilon_{jk} m_j m_k.
\end{align}
By considering the terms, $\varepsilon_{xx}$ and $\varepsilon_{xz}=\varepsilon_{zx}$, we can obtain
\begin{align}
[\vect{F}_\mathrm{MM}]_x=&b_1 \partial_x  m^2_x +b_2 \partial_Z m_x m_z
\sim-ikb_1m_2\sin2\phi_M=-ikb_1\sin2\phi_M \sum_{j}\psi_{\mathrm{M},j}(\vect{r})m_{j,2}(t),\nonumber\\
[\vect{F}_\mathrm{MM}]_z=&b_2\partial_x m_x m_z\sim -ikb_2m_1 \cos\phi_M=-ikb_2\cos\phi_M\sum_j\psi_{\mathrm{M},j}(\vect{r})m_{j,1}(t).
\end{align}
Thus, the magnon-phonon coupling term is rewritten as
\begin{align}
\frac{\int\mathrm{d}^3\vect{r}\vect{F}_\mathrm{MM}\cdot\vect{\psi}_A(\vect{r})}{\rho \int \mathrm{d}^3\vect{r} \vect{\psi}_A(\vect{r})\cdot\vect{\psi}_A(\vect{r})}=\sum_j\left[-kb_1\sin2\phi_M \frac{V_{\mathrm{OV,x},j}}{\rho V_\mathrm{A}}m_{j,2}(t)-kb_2\cos\phi_M \frac{V_{\mathrm{OV,z},j}}{\rho V_\mathrm{A}}m_{j,1}(t)\right].
\end{align}
By performing a rotating frame approximation and applying $\phi_M=0$, we obtain
\begin{align}
\dot{b}=\left(i\Delta-\frac{\kappa_A}{2}\right)b+i\sum_l g_{\mathrm{AM},l}m_{l,1}+b_\mathrm{in},
\end{align}
where $g_{\mathrm{AM},l}\equiv (b_2 k)/(2\omega_\mathrm{A}\rho) (V_{\mathrm{OV,z},l}/V_\mathrm{A})$ and $b_\mathrm{in}$ is the drive amplitude.

\section{Numerical analysis}
The magnon-phonon coupling rates were numerically estimated by solving the coupled mode equation in the frequency domain. Here, we took into account three thickness modes $\vect{m}_0$, $\vect{m}_1$, and $\vect{m}_2$, so the coupled mode equation had a 7x7 response matrix. By calculating the inverted response matrix and applying it to the driving field vector, $(b_\mathrm{in}, m_\mathrm{in,0},0,m_\mathrm{in,1},0,m_\mathrm{in,2},0)$, we finally obtained the phonon amplitude $b(\omega)$ with the driving field vector. The parameters, such as the saturation magnetic field, Gilbert damping, and etc., were optimized via a Monte-Carlo method with 10000 iterations. Here, we defined the cost function,
\begin{align}
\Psi=\sum_j\left|A_\mathrm{th}(H_{\mathrm{ext},j})-A_\mathrm{exp}(\mu_0 H_{\mathrm{ext},j})\right|^2
\end{align}
where $A_\mathrm{th}$ is the phonon amplitude numerically calculated with respect to the external magnetic field $H_{\mathrm{ext},j}$, and $A_\mathrm{exp}$ is the phonon amplitude experimentally measured from the phonon spectra by fitting with a Lorentz function for each $H_{\mathrm{ext},j}$. The optimized parameters are listed in Table I.
\begin{table}[hbt]
\centering
\begin{tabular}{cccc}\hline
   parameter &  notation & opt. values& unit\\ \hline
   Saturation magnetic field & $M_S$ & 0.19 & T\\
   Magneto-elastic coupling coefficient (shear)& $b_2$ & 3.74& T \\
   Gilbert damping & $\alpha$ & $1.21\times 10^{-4}$& none\\
   Exchange coefficient & $\beta$ &$2.67\times10^{-16}$& $\mathrm{m}^2$\\
   IDT window width &$\Delta_W$&$2.85 \kappa_\mathrm{M}$&none\\
   \hline
\end{tabular}
\end{table}
The error bars in Figs. 4(a) - 4(c) were calculated via the bootstrap method. First, multiple parameter sets were drawn from Gaussian distributions whose the mean value corresponded to the optimal parameter value and standard deviation corresponded to 2.5 \% of those values. Then, the residual sum of squares was recalculated.

\subsection{Contribution from the magnon excitation}
Here, we discuss the contribution from the magnon excitation, i.e., the finite value of $m_{\mathrm{in},0}$ and $m_{\mathrm{in},1}$. Figures S1(a) and (b) show the numerically calculated phonon spectra with and without mangnon excitation for the same parameters. As shown in Figs. 2(a) and 2(b), the phonon spectra calculated with magnon excitation are in good agreement with experimental results, whereas those without magnon excitation are not.

Moreover, we emphasize that the magnon excitation is coherent and not a thermal excitation. This contribution can be unveiled by taking account the phonon amplitude, 
\begin{align}
    b(\omega)=A^{-1}_{11}(\omega) b_\mathrm{in}+A^{-1}_{12}(\omega) m_{\mathrm{in},0}+A^{-1}_{13}(\omega) m_{\mathrm{in},1}\cdots .
\end{align}
The phonon spectra measured in our experiment is proportional to the square of the phonon amplitude,
\begin{align}
    |b(\omega)|^2=\left|A^{-1}_{11}(\omega) b_\mathrm{in}+A^{-1}_{12}(\omega) m_{\mathrm{in},0}+A^{-1}_{13}(\omega) m_{\mathrm{in},1}\cdots\right|^2.\label{response}
\end{align}
Importantly, Eq. (\ref{response}) includes the cross term when the magnon modes are coherently excited. Note that the numerical results in Fig. S1(a) show the case of coherent driving. Because these phonon spectra are in good agreement with the experimental results, we can conclude that the magnon was excited coherently and that it leads to phonon amplitude modulation via magnon-phonon interference. We also calculated the case with incoherent driving, as indicated by $\langle m_{\mathrm{in},j}(\omega) m_{\mathrm{in},k}(\omega')\rangle=\delta_{jk}\delta(\omega-\omega')$ [see Fig. S1(c)]. Although the phonon spectra shows the additional contribution from the magnon spectra, the results with no interference and a bright magnon branch obtained here differ significantly from the experimental results.

\begin{figure}[t]
	\begin{center}
		\vspace{-0.5cm}\hspace{-0.0cm}
		\includegraphics[width=18cm]{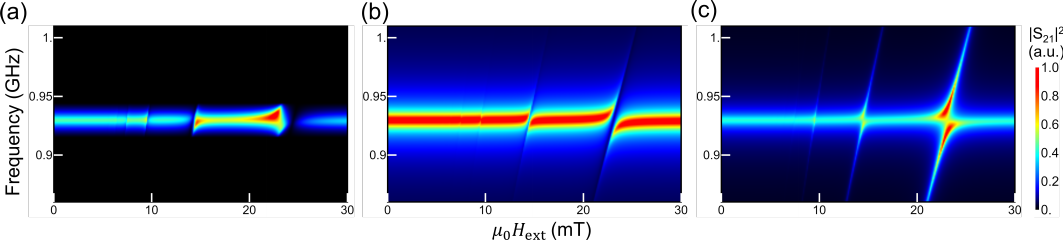}
		\vspace{0.5cm}
		\caption{Phonon response with respect to the external magnetic field (a) with coherent magnon excitation, (b) without magnon excitation, and (c) with incoherent magnon excitation. The experimental results match the results with the coherent magnon excitation shown in (a).}
		\label{fig S1}
		\vspace{-0cm}
	\end{center}
\end{figure}

\subsection{Excitation ratios in magnon-phonon interference}
The magnon modes are coherently excited via the rf field from the IDT electrodes in our magnomechanical system. From the numerical analysis, we can estimate the excitation ratio in the magnon-phonon interference from the observed phonon amplitudes. This ratio can be estimated by converting $b\to \sqrt{g_{\mathrm{AM},l}}b$, and $\vect{m}_l \to \sqrt{g_{\mathrm{MA},l}}\vect{m}_l$ in the coupled mode equation [Eqs. (S1) and (S2)]. Thus, the symmetrized equation of motion is given as
\begin{align}
\dot{\vect{m}_l}(t)=D_l\vect{m}_l(t)+\sum_{k\neq l}\tilde{R}_{k}\vect{m}_k+ig_{\mathrm{MA},l}(0,1)^T b(t)+\frac{\vect{m}_{\mathrm{in},l}}{\sqrt{g_{\mathrm{MA},l}}},\\
\dot{b}(t)=\left(i\Delta-\frac{\kappa_A}{2}\right)b(t)+ig_{\mathrm{AM},l} (0,1)^T\cdot \vect{m}_{l}(t)+\frac{b_\mathrm{in}}{\sqrt{g_{\mathrm{AM},l}}}.
\end{align}
where $\tilde{R}_k\equiv \sqrt{g_{\mathrm{MA},k}/g_{\mathrm{MA},l}}R_k$ denotes the rescaled dipole-dipole interaction matrix. Here, we define the excitation ratio by
\begin{align}
    \eta_l \equiv \sqrt{\frac{g_{\mathrm{AM},l}}{g_{\mathrm{MA},l}}}\frac{|\vect{m}_{\mathrm{in},l}|}{b_\mathrm{in}}=\sqrt{\frac{M_{S}}{2\gamma \rho\omega_A}}\frac{|\vect{m}_{\mathrm{in},l}|}{b_\mathrm{in}}.
\end{align}
From the numerical analysis, we obtain $\eta_0=5.84\pm0.70$, $\eta_1=2.50\pm0.25$, and $\eta_2=0.46\pm0.06$. 

\subsection{Effect of dipole-dipole interactions}
The dipole-dipole interaction between the different magnon thickness modes modulates the magnon eigenmodes, and that leads to additional phonon modulation via the magnon-phonon interaction. Here, a numerical calculation successfully reproduced the small phonon modulation appearing around the external magnetic field of 10 mT (see Fig. S2). The three small modulations in the phonon spectra observed in the experiment are also in the numerical results for the case with the dipole-dipole interaction whereas they disappear in the case without the dipole-dipole interaction. Thus, the dipole-dipole interaction between the thickness magnon modes plays a crucial role in modulating the phonon spectra via the magnon-phonon coupling.
\begin{figure}[t]
	\begin{center}
		\vspace{-0.5cm}\hspace{-0.0cm}
		\includegraphics[width=18cm]{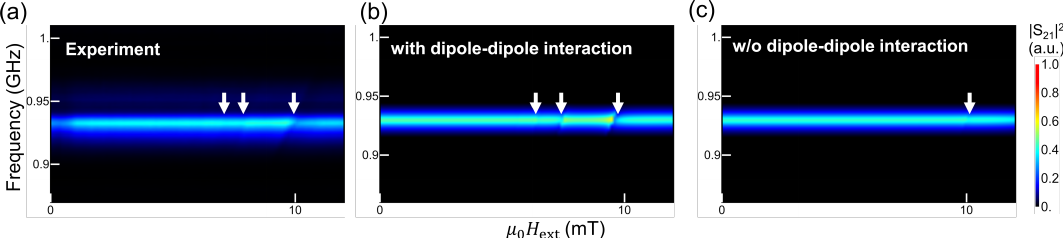}
		\vspace{0.5cm}
		\caption{Phonon response with respect to the external magnetic field around 10 mT (a) observed in the experiment, numerically calculated response with   the dipole-dipole interaction (b), and  numerically calculated response without the dipole-dipole interaction (c). The phonon spectra modulated via magnon-phonon coupling are highlighted by white arrows. The numerical results for the case with the dipole-dipole interaction are in good agreement in the experimental results in terms of the additional modulation in the phonon spectra.}
		\label{fig S2}
		\vspace{-0cm}
	\end{center}
\end{figure}

\end{document}